\documentclass[12pt]{article}
\usepackage{cite}
\usepackage{amsmath,amssymb,amsfonts}
\usepackage{algorithmic}
\usepackage{graphicx}
\usepackage{textcomp}
\usepackage{hyperref}
\usepackage{booktabs}
\usepackage{authblk}

\usepackage{gensymb}
\def\code#1{\texttt{#1}}
\def\BibTeX{{\rm B\kern-.05em{\sc i\kern-.025em b}\kern-.08em
    T\kern-.1667em\lower.7ex\hbox{E}\kern-.125emX}}

\title{Empirical Measurements of AI Training Power Demand on a GPU-Accelerated Node}

\title{Empirical Measurements of AI Training Power Demand on a GPU-Accelerated Node}

\author[1]{Imran Latif}
\author[2]{Alex C. Newkirk}
\author[1]{Matthew R. Carbone}
\author[3]{Arslan Munir}
\author[1]{Yuewei Lin}
\author[4]{Jonathan Koomey}
\author[1]{Xi Yu}
\author[1]{Zhihua Dong}
\affil[1]{Computing and Data Sciences Directorate, Brookhaven National Laboratory, Upton, NY, USA}
\affil[2]{Building Technology and Urban Systems, Lawrence Berkeley National Laboratory, Berkeley, CA, USA}
\affil[3]{Department of Electrical Engineering and Computer Science, Florida Atlantic University, Boca Raton, FL, USA}
\affil[4]{Koomey Analytics, San Francisco, CA, USA}

\begin{document}

\maketitle

\begin{abstract}
The expansion of artificial intelligence (AI) applications has driven substantial investment in computational infrastructure, especially by cloud computing providers. Quantifying the energy footprint of this infrastructure requires models parameterized by the power demand of AI hardware during training. We empirically measured the instantaneous power draw of an 8-GPU NVIDIA H100 HGX node during the training of open-source image classifier (ResNet) and large-language models (Llama2-13b). The maximum observed power draw was approximately 8.4~kW, 18\% lower than the manufacturer-rated 10.2~kW, even with GPUs near full utilization. Holding model architecture constant, increasing batch size from 512 to 4096 images for ResNet reduced total training energy consumption by a factor of 4. These findings can inform capacity planning for data center operators and energy use estimates by researchers. Future work will investigate the impact of cooling technology and carbon-aware scheduling on AI workload energy consumption.
\end{abstract}

Keywords: AI training, Sustainable computing, GPU power measurements

\section{Introduction}
The modern era of deep learning required both algorithmic innovation and dedicated hardware. As researchers identified scaling relationships in artificial intelligence (AI) applications between model scale, total training, and performance~\cite{kaplan_scaling_2020}, computational hardware optimized for these workloads enabled the training of enormous models with tens-to-hundreds of billions of parameters, using thousands of graphics processing units (GPUs). Beginning in late 2022, these steadily improving capabilities captured consumer interest. Cloud infrastructure providers responded with dramatic investment in hardware~\cite{rattner_ai_2024}, both to meet this demand and in anticipation of further growth. 

While specialized AI chips such as GPUs or tensor processing units are energy efficient on a per-operation basis, the expansion of computational infrastructure for AI applications necessarily carries a substantial energy and environmental footprint. While organizations training and deploying AI know their electricity consumption, any projections of future demand must model future load. Additionally, any actors without direct access to the relevant hardware, including electrical grid operators making capacity planning decision or researchers assessing sustainability, rely on these models. Energy-use models in turn rely on a parameterized assumption of the power demand of hardware during AI training. To support these efforts to model energy use, we have empirically measured the instantaneous power demand of current generation AI hardware across a variety of AI training workloads, and assessed the impact of model architecture and memory-batch size on energy performance. 

This work is structured as follows. In Sec.~\ref{sec:background and motivations}, we begin with background on the growth of AI infrastructure and the need for accurate power measurements to inform energy use models. Sec.~\ref{sec:methods} details our empirical methodology, including the hardware configuration (Sec.~\ref{sec:experimental setup and equipment}) tested, the AI training workloads used, and the power measurement approach. We then present and discuss the empirical power demand recorded across three training runs, varying in architecture and memory batching, in Sec.~\ref{sec:results} and \ref{sec:discussion}. Finally, we conclude with a discussion of the limitations of this work, the implications of these findings for data center capacity planning and energy estimation, and directions for future research. 

\subsection{Significant results} 
The most significant results of our work are summarized as follows:
\begin{itemize}
\item The maximum observed power draw during a GPU/CPU burn control was 8.4~kW, well below the 10.2~kW manufacturer rated maximum;
\item Llama training GPU load was 93\% on average, with a median power draw of 7.9~kW, indicating, \textit{also} well below the rated maximum, which highlights the need for empirical data for data center energy use estimation;
\item Increasing ResNet training batch size from 512 to 4096 in image classifier training used 1~kW higher power on average but 4$\times$ less energy.
\end{itemize}

\section{Background and Motivations} \label{sec:background and motivations}
The commercialization of AI and the substantial hardware investment it requires have drawn increased attention to the energy footprint of these workloads. Coinciding with heightened scrutiny from scholars and the public, companies have reduced the amount of data they release on hardware and metered energy~\cite{alben_computing_2024}. This shift is partially due to the maturation of AI applications. As AI transitioned from an academic curiosity to a commercialized product, hardware operation details became a potential source of competitive advantage. Researchers internal to firms conducting training already have access to relevant aggregate data, e.g., their power bill. Several estimates of training energy intensity rely on this proprietary internal power draw data~\cite{patel_characterizing_2024, patterson_carbon_2022, wu_sustainable_2022}. Academic researchers instead have to rely on what data they have available, with one common approach to multiply the manufacturer rated power of AI-optimized GPUs or servers~\cite{luccioni_estimating_2023} by the reported training time (typically in GPU-hours) of a given model~\cite{de_vries_growing_2023, lacoste_quantifying_2019, faiz_llmcarbon_2023}. While this provides valuable bounds, it does not enable granular estimates of node-level power draw, assuming either full utilization of all non-GPU components or neglecting them entirely.

GPU component-level utilization is well specified, in part because most AI-optimized hardware reports it through data center infrastructure management (DCIM) software. In contrast, node-level energy use is often approximated~\cite{bouza_how_2023}. There is research grounded in empirical metering of all components or entire nodes~\cite{hodak_towards_2019, wang_energy_2023}, but these are not generally performed on current generation hardware. There is also some ongoing work characterizing the distinct load profiles of physical modeling and AI workloads~\cite{govind_comparing_2023, shankar_trends_2022, zhao_power_2023}, though they typically focus on the chip/processor rather than node. Calibrating these estimates with node-level measured power serves to shed further light on the energy use of AI. While AI training and performance benchmarks have been proposed and implemented in the industry~\cite{farrell_mlperf_2021}, they have not kept pace with commercial system scale-up~\cite{alben_computing_2024}. Robust empirical evidence should reflect recent hardware improvements and infrastructure investments. Additionally, there is an opportunity to compare traditional simulation workloads against AI training, as the two are performed on increasingly overlapping hardware configurations.

\section{Methodology} \label{sec:methods}
To address this need for empirical measurement of in-workload training power draw on contemporary ML specialized nodes hardware, we conducted a series of training runs on an 8-GPU Nvidia H100 HGX Node at the Brookhaven National Laboratory Scientific Data and Computing Center. 

We aimed to characterize the power profiles of a diverse set of AI workloads. To ensure replicability and relevance to real-world applications, we selected open-source models and datasets widely used in the ML community, such as by MLPerf. While proprietary models and datasets used in production settings may differ in specifics, open-source workloads that are commonly used for benchmarking, testing, and research provide the best available approximation. As the computational intensity differs between architecture, our chosen workloads included both an image network: ResNet, and a modern large language model: Llama2-13b.

To further contextualize these results beyond our chosen ML training workloads, we also performed an hardware stress test used for benchmarking known as a ``GPU burn". We selected Llama2-13b to approximate a production workload which would fully utilize available computation. To ensure the validity of this approach, we also conducted a hardware stress test with GPU, CPU and memory utilization maximized. This stress test provides an upper bound on the power draw of the system under maximum load, serving as a control experiment for the ML workload measurements. By comparing the power demand from this stress test, we could evaluate the applicability of our production workloads to a theoretical upper bound case for single-node power demand. 

The following subsections detail our experimental setup, including the specific hardware configuration, software environment, model architectures, and datasets used, our methodology for data collection and analysis, and limitations of our approach.
 
\section{Experimental Setup and Equipment} \label{sec:experimental setup and equipment}

\subsection{Computational Hardware}

\subsubsection{Cooling and Power Infrastructure}
The cooling of the HGX node in the standard 19" wide, 42U rack is done by Rear-Door Heat Exchangers (RDHx). The RDHx is supplied with 15.5 \textdegree C facility chilled water supply to maintain 23.8 \textdegree C ambient cooling temperature. The power to the HGX node is provided via rack mounted power distribution units (PDU's) using six standard C13 outlets. The 208~V/3-phase, 50~A PDU's are connected to overhead bus taps with twist lock mechanism. The power to the RDHx unit is also supplied by the Rack PDU's, however this would be coming from the other PDU mounted in the same computer rack. 

\subsubsection{CPU and GPU Information}
CPU model: AMD EPYC 9354 32-Core Processor with 1.5~TB memory. GPU model: NVIDIA H100 with 80~GB memory. All the experiments are conducted with single node with 8 H100 GPUs.

\subsection{Training workload and Software Configuration}
We conducted two types of experiments: image classification tasks and visual question answering (VQA) tasks, to measure the power consumption of the models during training.

\subsubsection{Image Classification}

For the image classification experiments, we used the ResNet-152~\cite{he2016deep} architecture. The CIFAR-10~\cite{Krizhevsky2009LearningML} dataset was employed for these tasks. It consists of 60k $32\times32$ color images spread across 10 categories (airplane, automobile, bird, cat, deer, dog, frog, horse, ship, and truck), with 50k images for training and 10k for testing.

\subsubsection{VQA} 

In the VQA experiments, we utilized a Multi-Modal Large Language Model LLaVA~\cite{liu2024improved} that combines a visual encoder, CLIP-L-14~\cite{radford2021learning}, with a language model decoder, Llama2-13b~\cite{touvron2023llama}. The visual encoder processes images, while the Llama2 model, with $\sim$13 billion parameters, focuses on understanding and generating answers based on text. The VQA tasks were trained using a combined dataset of 665k pairs, including image-caption and image-question-answer pairs. These pairs were sourced from COCO~\cite{lin2014microsoft}, GQA~\cite{hudson2019gqa}, OCR-VQA~\cite{mishra2019ocr}, TextVQA~\cite{singh2019towards}, and Visual Genome~\cite{krishna2017visual}, each providing rich visual-textual alignment. The COCO~\cite{lin2014microsoft} dataset comprises approximately 118k training images, 5k validation images, and 40k test images, with resolutions ranging from $480\times640$ to $1024\times768$ pixels, and is primarily used for object detection, segmentation, and image captioning tasks. GQA~\cite{hudson2019gqa} includes around 72k images paired with 943k questions, focusing on visual question answering with image sizes typically around $480\times640$ pixels. OCR-VQA~\cite{mishra2019ocr} contains 50k images and questions, designed for text recognition within images. TextVQA~\cite{singh2019towards} offers 28,408 images and 45,336 questions, requiring models to interpret text embedded in visual content. Visual Genome~\cite{krishna2017visual} provides 108k images, annotated with objects, attributes, and relationships, and is used for tasks like object detection, scene graph generation, and visual reasoning, with image sizes generally between $480\times640$ and $1024\times768$ pixels.

\subsubsection{Implementation Details}
All tasks are implemented using Distributed Data Parallel (DDP) in PyTorch 2.2. For the image classification task, we train the ResNet model from scratch with a batch size of 512 and 4096 for 200 epochs, optimizing the entire model using Stochastic Gradient Descent (SGD) with an initial learning rate of 0.1, a momentum of 0.9, and a weight decay of $5 \times 10^{-4}$. We employ a cosine annealing learning rate scheduler during training, which adjusts the learning rate based on a cosine function to enhance training performance and convergence. For the Visual Question Answering (VQA) task, we fine-tune the LLaVA~\cite{liu2024improved} model on a combined dataset. The LLaVA~\cite{liu2024improved} model comprises a visual encoder, a projector, and a large language model (LLM). Specifically, we use CLIP-L-14~\cite{radford2021learning} as the visual encoder, a two-layer MLP with GeLU activation as the projector, and Llama2-13b~\cite{touvron2023llama} as the LLM. During training, we keep the visual encoder and projector fixed and train only the LLM. The batch size is set to 128 with learning rate of $2 \times 10^{-5}$ and warm up ratio is 0.03 with cosine learning rate scheduler.

\subsubsection{GPU Burn}
To evaluate the external validity and real-world applicability of our experimental results, we utilized the open-source GPU Burn tool~\cite{timonen_wiliccgpu-burn_2024}. A GPU burn is designed to stress test multi-GPU setups by maximizing GPU memory utilization and compute load while validating calculation correctness. The tool forks a process for each GPU, allocates a high percentage of available memory, and performs continuous matrix multiplications using the CUBLAS library. We obtained the gpu-burn source code from the official GitHub repository at \code{github.com/wilicc/gpu-burn} and compiled it on our target Linux system following the provided instructions.

For our specific test scenario, we configured gpu-burn to utilize the default memory setting which is at 90\% (72~GB) of available GPU memory (80~GB/GPU). Additionally, we enabled tensor core usage via the \code{-tc} flag to leverage the full computational capabilities of the GPUs and simulate peak compute demand. GPU Burn was executed in two configurations: first, as a standalone GPU stress test, and second, in parallel with the stress-ng CPU stress testing tool to assess performance under simultaneous CPU and GPU load. stress-ng, available at \code{github.com/ColinIanKing/stress-ng}, is a workload generation tool used for stress testing and benchmarking various components of a computer system, including CPUs, memory, disk I/O, and more. It provides a wide range of stress tests, allowing users to simulate different types of workloads to evaluate system performance and stability. The stress-ng code was ran with 256 tasks for 3D  matrix operations. The run used 1.3~TB of 1.5~TB available host memory.

The combination of maximizing memory utilization, enabling tensor cores, and introducing CPU stress alongside the GPU burn aims to characterize the system's behavior under maximum-case computational demand.

\subsection{Power Measurement}
The GPU power draw was measured using the Weights \& Biases (wandb)~\cite{wandb} developer interface. The DCIM, provided by NLYTE software, is used to manage, monitor and optimize the data center's power and cooling infrastructure. Power and current data from the rack mounted PDU is pulled into the DCIM system over the network using communication protocols like SNMP and Modbus. The DCIM collects and processes this data, allowing operators to monitor, analyze, and optimize power usage at the rack level. The DCIM reports real-time data, including metrics like power draw, and alerts for potential issues related to critical infrastructure in the data center.

DCIM software collects the available telemetry on the PDU and can alarm on specified metrics or on SNMP traps. Telemetry is collected from items such as, but not limited to power, current, voltage (L2L and L2N), IT devices, servers, PDU’s, transfer switches, UPS's, switch gear, transformers, branch circuit monitoring, and other end point devices.

\section{Results} \label{sec:results}
This section reports the observed ranges of power draws, examines the impact of factors like model architecture and batch size on energy usage, and analyzes the relationship between GPU utilization and power demand. These results offer insights into the power requirements and energy footprint of AI training on contemporary hardware. 


\begin{figure}[htbp!]
    \centering
    \includegraphics[width=1.0\columnwidth]{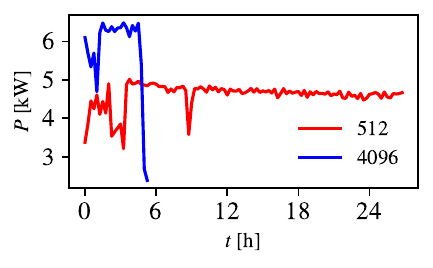}
    \caption{Instantaneous pre-training power demand measured in kW. Data corresponds to training ResNet with a batch size of 512 and 4096 images. Total energy usage for these two cases were computed by integrating the curves and are 123 and 30 kWh, respectively.}
    \label{fig:power demand resnet}
\end{figure}

The training of the ResNet image classifier with a batch size of 512 images took over 26 hours. The median node power demand was 4.68~kW, with a maximum power demand measured during this training of 5.02~W. This was followed by the training of ResNet with a batch size of 4096 images, which completed in 5 hours. The node demanded a median power draw of 6.26~kW during this workload, with a maximum measured demand of 6.48~kW.The instantaneous power demand for these experiments is showcased in Fig.~\ref{fig:power demand resnet}.

\begin{figure}[htbp!]
    \centering
    \includegraphics[width=1.0\columnwidth]{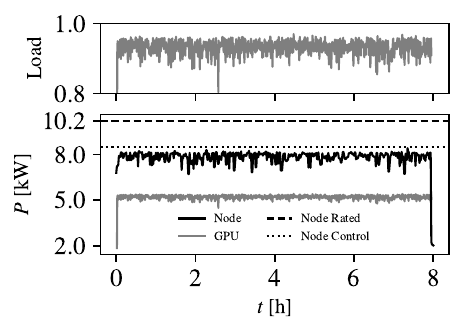}
    \caption{Instantaneous pre-training power demand measured in kW, and average total GPU load, for the Llama2-13b parameter model. The rated power of the system (10.2~kW) is shown as a horizontal dashed line. The median node power draw from the GPU$+$CPU burn control experiment is shown as a horizontal dotted line (at 8.43~kW).\label{fig:power demand llama}}
\end{figure}

The instantaneous power demand results for training Llama2-13b are shown in Fig.~\ref{fig:power demand llama}. Llama2-13b was trained with a batch size of 128 image question-answer pairs, completing in 8 hours. The median node power consumption was 7.92~kW, with a maximum measured power demand of 8.42~kW. GPU-level utilization during training (shown as the ``GPU" label in Fig.~\ref{fig:power demand llama}) was close to maximum for all 8 chips, suggesting near computational saturation. It is noteworthy that our empirically measured power draw never approached the manufacturer rated maximum of 10.2~kW (``Node Rated").

The dotted line in Fig.~\ref{fig:power demand llama} (``Node Control") shows the total node power consumption during the maximally intensive GPU/CPU Burn stress tests, which maximized available GPU memory and tensor core operations. For this workload, where the GPU and CPU stress tests ran concurrently, the median power demand was 8.43~kW (with a max of 8.48~kW), only $\sim$400~W greater than the median demand of the Llama workload. The minimal difference in power consumption between the GPU stress test and the real-world Llama training workload demonstrates that our empirical measurements offer a reliable characterization of the power requirements for computationally demanding workloads on this hardware.


\begin{table*}[htbp]
\centering
\small 
\setlength{\tabcolsep}{4pt} 
\caption{Power sampled at 1-minute intervals over training duration. Batch size in images/tokens for ResNet/Llama models. \label{table:training_workload_optimized}}
\begin{tabular}{lcccccr}
\toprule
Training & Model & Median & Batch & Duration & Total & Avg. GPU \\
Workload & Task & Power (kW) & Size & (min) & Energy (kWh) & Util. (\%) \\
\midrule
Idle Node & N/A & 1.86 & N/A & N/A & N/A & N/A \\
Resnet-512 & Image & 4.68 & 512 & 1,605 & 123 & 36 \\
Resnet-4096 & Image & 6.26 & 4,096 & 315 & 30 & 77 \\
Llama2-13b & VQA & 7.92 & 128 & 480 & 62 & 93 \\
GPU Burn (no CPU) & Stress & 7.97 & N/A & 60 & 8 & 100 \\
GPU Burn (+CPU) & Stress & 8.43 & N/A & 77 & 10.5 & 100 \\ 
\bottomrule
\end{tabular}
\end{table*}

\section{Discussion} \label{sec:discussion}

The power demand measurements in this study are specific to the particular combination of IT and cooling hardware in the tested system configuration, which may limit the generalizability of the findings. GPUs are constant within a product specification, but the supporting memory, supervisory CPUs, interconnect, and fan components vary between vendors. Accordingly, these results may be confounded by specific configuration of this node. Additionally, while components generally operate withing their specified bounds, the material variability of manufacturing logic devices means the precise power performance varies slightly chip-to-chip. Future work should extend these empirical results to include a greater variety of server configurations.

The total energy use and instantaneous power demand of a given training workload were sensitive to hyperparameter choice. The training task was held constant between ResNet-1 and ResNet-2, they only varied by batch size. The smaller batch size in ResNet-1 resulted in a 900~W decrease in node power demand, but increased total IT energy consumed by 90~kWh. When considering the total energy footprint of a given workload, IT power demand, runtime, and cooling efficiency all contribute. 

Owing to the asymmetric loss functions for over- vs. under-provision of power for computational hardware, manufacturer-rated power limits typically serves as a design guideline for infrastructure procurement. The rack-level power distribution at the Scientific Data and Computing Center supplies 70~kW of power to the racks which contain 5 Nvidia HGX nodes. These racks were designed according to the manufacturer rated maximum, with a margin of redundant power of approximately 27\%. These results suggest that existing infrastructure could support an additional 8-GPU node while maintaining design-specified redundancy. 

All training runs were conducted on a single node system, there is an opportunity for future work to extend these analyses to multi-nodal training. Additionally, distributed training across multiple data centers should also be evaluated as this might expand opportunities for carbon-aware load shifting and scheduling, but could also potentially increase environmental footprint if data transmission or an increase in total training energy exceed associated carbon savings. Full visibility into the component-level share of node power draw is another opportunity for future research efforts. While technically challenging, detailed sub-metering of all active components in the node would identify the sensitivities of instantaneous power draw within different workloads, enabling hardware optimization. 

These power data were collected using a single configuration of IT and cooling hardware. The active components within AI training servers include GPU's, supervisory CPU's, memory, storage drives, interconnect, and fans. While holding the active components of a given server constant, there may be interaction effects between the IT and cooling systems which could affect workload power demand. More effective cooling technologies, such as direct-to-chip liquid cooling, could allow the same computational utilization with lower fan usage. This would in practice be a shift of a greater share of the cooling services onto the more efficient system, enabling the same computational intensity at a lower system energy use. Future work should evaluate impact of cooling technology on node-level power demand, as well as other configuration, hyper-parametric, and infrastructure determinants of workload energy use. 

\section{Conclusion} \label{sec:conclusion}
The increase in AI related computational demand has been enabled by investment in hardware infrastructure. This expansion has also prompted greater scrutiny into the energy footprint of this hardware. Estimates of AI energy use depend on parameterized value for in-workload power draw. As hardware has evolved, empirical measurement of this power demand has not kept pace. To address this, we have measured the instantaneous power demand of a variety of AI and ML workloads on can 8 NVIDIA H100 HGX Node. 

We found across all workloads, a maximum power demand of 8.48~kW, as compared to a manufacturer rated maximum power of 10.2~kW. This result was consistent with GPU utilization time-series data, which showed all 8 GPU's at or near maximum utilization during these time intervals. We also found that the total energy use of a given workload depends on memory batch size, holding hardware constant, with an adjustment in memory batch size for training an image classifier reduced average instantaneous power demand by 1~kW while increasing total pre-training energy use by a factor of 4. The findings from this research may assist data center owners and operators in optimizing infrastructure provisioning and enable them to deploy additional IT hardware using the same power distribution. Future work will investigate the impact of cooling technology on the energy use of a given workload, and evaluate the potential impact of carbon-aware scheduling on AI training workloads.

\bibliographystyle{plain}
\bibliography{bib}

\end{document}